\documentclass[twocolumn,superscriptaddress,pr,10pt]{revtex4-1}
\usepackage{verbatim}
\usepackage{braket}
\usepackage{amsmath,amssymb}
\usepackage{tikz}
\usepackage{graphicx}
\usepackage{color}
\usepackage[colorlinks,bookmarks=false,citecolor=blue,linkcolor=red,urlcolor=blue]{hyperref}
\usepackage{times}

\begin{document}

\title{Out-of-equilibrium protocol for R\'enyi entropies via the Jarzynski equality}

\author{Vincenzo Alba}
\affiliation{International School for Advanced Studies (SISSA),
Via Bonomea 265, 34136, Trieste, Italy, 
INFN, Sezione di Trieste}

\date{\today}

\begin{abstract} 
In recent years entanglement measures, such as the von Neumann and the R\'enyi entropies, 
provided a unique opportunity to access elusive feature of quantum many-body systems. 
However, extracting entanglement properties analytically, experimentally, or in numerical 
simulations can be a formidable task. Here, by combining the replica trick and the Jarzynski 
equality we devise a new effective {\it out-of-equilibrium} protocol for measuring the 
equilibrium R\'enyi entropies. The key idea is to perform a quench in the geometry of the replicas. 
The R\'enyi entropies are obtained as the exponential average of the work performed during 
the quench. We illustrate an application of the method in classical Monte Carlo simulations, 
although it could be useful in different contexts, such as in quantum Monte Carlo, or 
experimentally in cold-atom systems. The method is most effective in the quasi-static 
regime, i.e., for a slow quench. As a benchmark, we compute the R\'enyi 
entropies in the Ising universality class in $1$$+$$1$ dimensions. We find 
perfect agreement with the well-known Conformal Field Theory (CFT) predictions. 
\end{abstract}


\maketitle

\section{Introduction}

\begin{figure}[t]
\includegraphics*[width=0.93\linewidth]{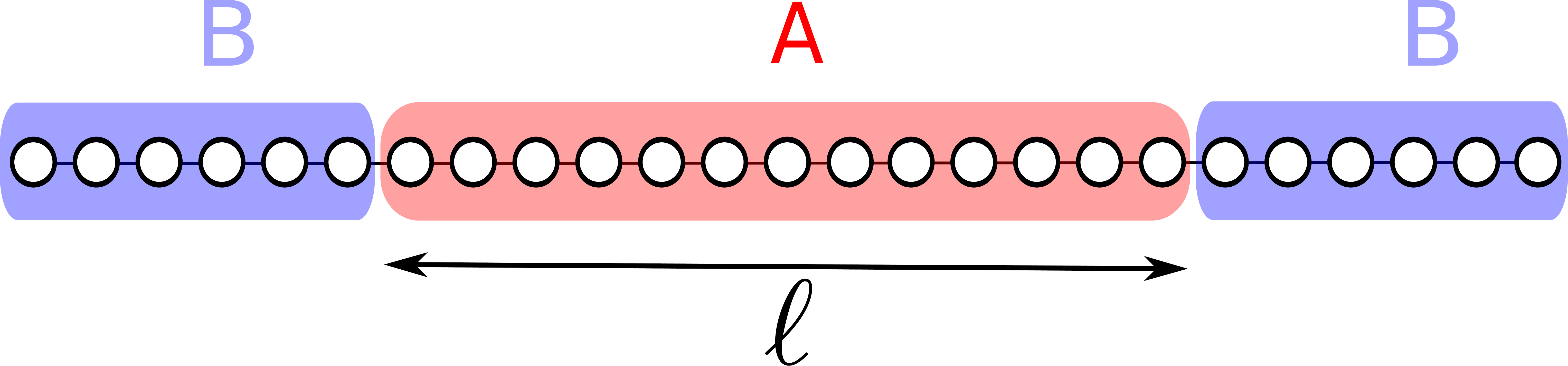}
\caption{ The bipartition used in this work. A chain of length $L$ with periodic 
 boundary conditions is divided into part $A$ of length $\ell$ and its complement 
 $B$. Here we are interested in the R\'enyi entropy $S^{{}_{(n)}}_A$  of $A$. 
}
\label{fig0}
\end{figure}

In recent years entanglement measures have arisen as new diagnostic tools to unveil 
universal behaviors in quantum many-body systems. Arguably, the most popular and 
useful ones are the R\'enyi entropies and the von Neumann entropy~\cite{amico-2008,
calabrese-2009,eisert-2010,laflorencie-2016} (entanglement 
entropies). Given a system in a pure state $|\psi\rangle$ and a bipartition into an 
interval $A$ and its complement (see Figure~\ref{fig0}), the R\'enyi entropies 
$S_A^{{}_{(n)}}$ for part $A$ are defined as 
\begin{equation}
\label{renyi}
S_A^{(n)}\equiv-\frac{1}{n-1}\ln\textrm{Tr}\rho_A^n, 
\end{equation}
with $\rho_A\equiv\textrm{Tr}_B|\psi\rangle\langle\psi|$ the reduced density 
matrix of $A$, and $\textrm{Tr}\rho_A^n$ its $n$-th moment. The limit $n\to 1$ 
defines the von Nemann entropy $S_A\equiv-\textrm{Tr}\rho_A\ln\rho_A$. 
Due to $\rho_A$ being non-local, extracting $S_A^{n()}$ 
analytically, experimentally, or even in numerical simulations can be a challenging task, 
execpt for free-fermion and free-boson models, for which the entropies 
can be obtained exactly in arbitrary dimensions~\cite{eisler-2009}. 

\begin{figure}[t]
\includegraphics*[width=0.99\linewidth]{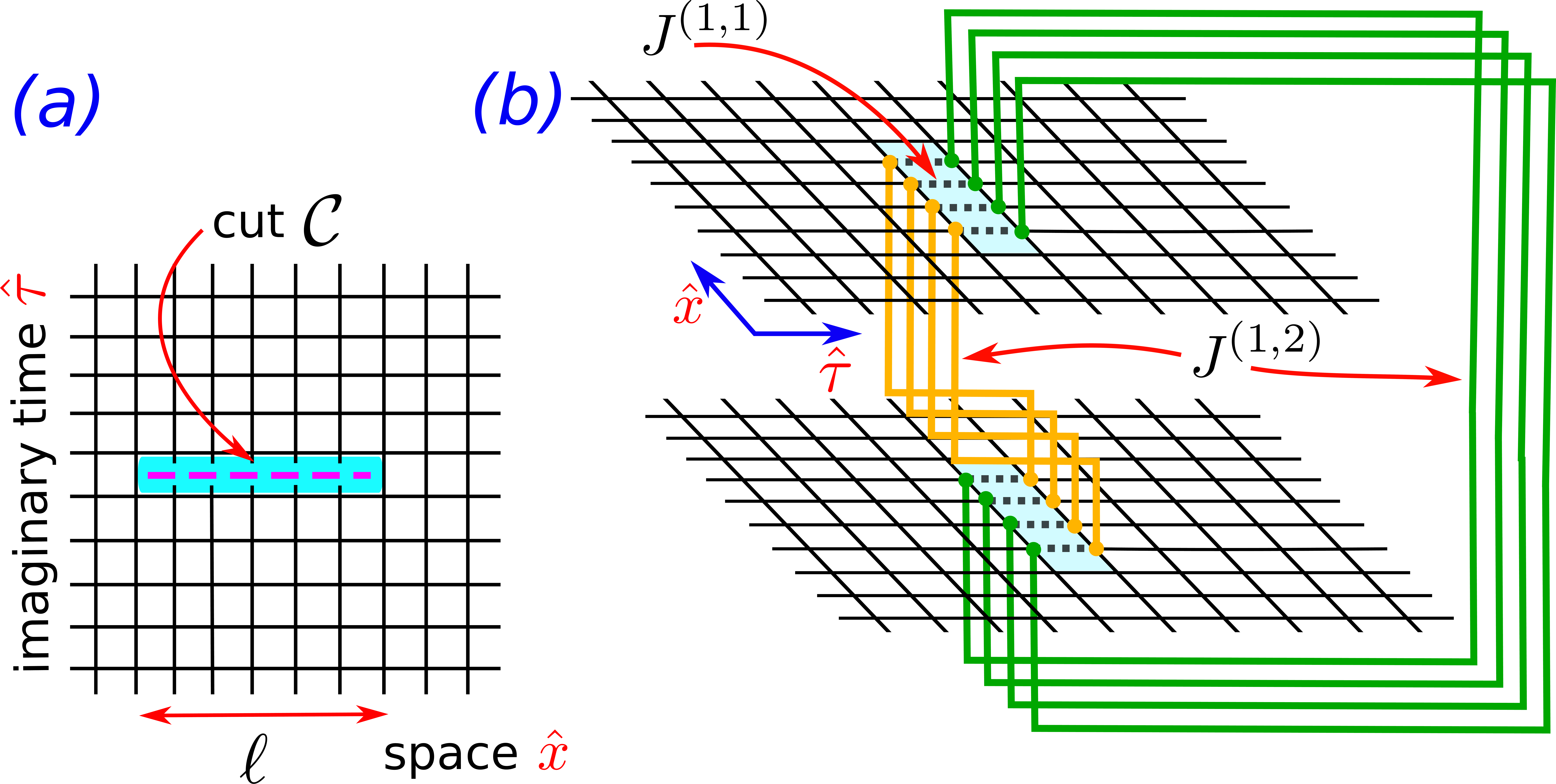}
\caption{ Replica geometry for calculating the R\'enyi entropy 
 $S^{{}_{(2)}}_A$. (a) The single sheet. The vertical and horizontal directions 
 are the imaginary time ($\hat\tau$) and spatial ($\hat x$) 
 directions, respectively. Periodic boundary conditions in both 
 directions are used. The horizontal dashed line denotes the branch cut 
 ${\mathcal C}$ lying on subsystem $A$ (see Figure~\ref{fig1}). 
 (b) Two sheets joined through 
 the cut (shaded regions). Spins around the cut and on different replicas 
 interact with coupling $J^{\scriptscriptstyle (1,2)}$. Spins around the cut and on 
 the same replicas interact with couplings $J^{\scriptscriptstyle (1,1)}$ and 
 $J^{\scriptscriptstyle (2,2)}$ (dotted links). The $2$-sheets 
 Riemann surface corresponds to $J^{\scriptscriptstyle (1,1)}=J^{\scriptscriptstyle 
 (2,2)}=0$ and $J^{\scriptscriptstyle (1,2)}=1$. The couplings $J^{{
 \scriptscriptstyle{(1,2)}}}$ are quenched during the simulation. 
}
\label{fig1}
\end{figure}

A large class of effective measurement protocols for the R\'enyi entropies 
are based on the {\it replica trick}. The key observation is that 
for a generic model $\textrm{Tr}\rho_A^n$ can be obtained as~\cite{calabrese-2004} 
\begin{equation}
\label{rep-trick}
\textrm{Tr}\rho_A^n=\frac{{\mathcal Z}_n(A)}{{\mathcal Z}^n}. 
\end{equation}
where ${\mathcal Z}\equiv\textrm{Tr}e^{-\beta{\mathcal H}}$, and ${\mathcal Z}_n(A)$ is 
the partition function on the so-called $n$-sheets Riemann surface (see Figure~\ref{fig1} 
(b)), which is defined by ``gluing'' together $n$ independent replicas of the model through 
part $A$. Importantly, the replica trick lies at the heart of all the known methods for measuring 
entanglement in cold-atom experiments~\cite{cardy-2011,abanin-2012,daley-2012,islam-2015,
kaufman-2016,pichler-2016}. For instance, very recently (see Ref.~\onlinecite{islam-2015}) 
$S_A^{{}_{(2)}}$ has been succesfully measured in ultra-cold bosonic systems from the  
interference between two replicas of a many-body state. 
Moreover, the ratio of partition functions in~\eqref{rep-trick} can be sampled using 
classical~\cite{buividovich-2008,caraglio-2008,alba-2010,gliozzi-2010,alba-2011,alba-2013} and 
quantum Monte Carlo techniques~\cite{hastings-2010,melko-2010,singh-2011,isakov-2011,humeniuk-2012,
kaul-2013,inglis-2013,iaconis-2013}, providing an efficient method to calculate 
R\'enyi entropies. Extensions of these techniques for systems in the 
continuum~\cite{herdman-2014}, and interacting fermions~\cite{zhang-2011,tubman-2012,mcminis-2013,grover-2013,assaad-2014,
broecker-2014,wang-2014,shao-2015,drut-2015,drut-2016,porter-2016} are also available. Monte Carlo methods 
work effectively in any dimension and at any temperature, for sign-problem-free models. 
Oppositely, the Density Matrix Renormalization Group~\cite{uli-2005,uli-2011} (DMRG) 
provides the most effective way to access the full spectrum of $\rho_A$ for one-dimensional systems, 
whereas is less effective in higher dimensions. 
A severe issue of all the replica-trick-based protocols is that as the size $\ell$ of $A$ increases, 
${\mathcal Z}_n(A)/{\mathcal Z}^n$ is dominated by rare configurations. The commonly used 
strategy to mitigate this issue is based on the {\it increment trick}~\cite{hastings-2010,
alba-2011,humeniuk-2012}. This consists in splitting the ratio in~\eqref{rep-trick} as a product 
of intermediate terms, which have to be measured separately. Their number 
typically grows as $\ell$, which is the main drawback of the protocol. 
%

Here we propose a new {\it out-of-equilibrium} framework for measuring the R\'enyi entropies.  
Similar protocols to access entanglement in cold-atom 
experiments have been explored in Ref.~\onlinecite{cardy-2011} and Ref.~\onlinecite{abanin-2012}. 
Our approach combines the replica trick~\eqref{rep-trick} and the Jarzynski equality~\cite{jarzynski-1997}. 
Crucially, the latter allows to relate the ratio of partition functions corresponding to two 
equilibrium thermodynamic states to the exponential average of the work performed during an {\it 
arbitrary} far-from-equilibrium process connecting them. The idea of the method is to modify 
(i.e., quenching) the geometry of the replicas, gradually driving the system from the geometry 
with $n$ independent replicas to that of the $n$-sheets Riemann surface. Using the Jarzynski 
equality, the R\'enyi entropies are then extracted from the statistics of the work performed during 
the quench, similar to Ref.~\onlinecite{cardy-2011}. Some applications of the Jarzyinski 
equality to detect entanglement have been also presented 
in Ref.~\onlinecite{hide-2010}. The efficiency of the method 
depends dramatically on the rate $\theta$ at which the geometry is 
modified. Precisely, the number of independent protocol realizations needed 
increases upon increasing $\theta$. However, in the 
quasi-static regime, i.e., large $\theta$, $S_A^{{}_{(n)}}$ can be 
extracted from a single realization. In this regime the R\'enyi entropies depend 
only on the average work and the standard deviation of the work fluctuations, reflecting 
that the work distribution function becomes gaussian. 

Here we illustrate the approach in the framework of classical Monte Carlo simulations. 
Specifically, we focus on the Ising universality 
class in $1$$+$$1$ dimensions, although the method works in any dimension, and it could be 
extended to quantum Monte Carlo. Remarkably, even for moderately large system 
sizes the R\'enyi entropies can be extracted in a single simulation, in contrast 
with methods using the increment trick, which require typically $\sim\ell$ independent 
simulations. 

\section{Method: Quenching the replica geometry} 
The replica geometries used in the method are illustrated in Figure~\ref{fig1}. Panel~\ref{fig1} 
(a) shows a single replica (sheet), where $\hat x$ and $\hat\tau$ are interpreted as the spatial and 
imaginary time directions, respectively. The single-sheet partition function ${\mathcal Z}$ is obtained 
as the path integral ${\mathcal Z}\equiv\int {\mathcal D}[\phi]e^{-{\mathcal S}(\{\phi\})}$, where 
$\phi(x,\tau)$ is a field and ${\mathcal S}$ is the euclidean action of the model. We impose the 
periodic boundary conditions $\phi(x,\tau)=\phi(x,\tau+L_\tau)$ and $\phi(x,\tau)=\phi(x+L_x,\tau)$, 
with $L_x$ and $L_\tau$ the number of sites along the $\hat x$ and $\hat\tau$ direction, 
respectively. For simplicity, we assume interaction only between nearest-neighbor 
sites $\langle i,j\rangle$, i.e., $S=\sum_{\langle i,j\rangle}F(\phi_i,\phi_j)$, with $F$ 
the interaction strength. In the replica trick~\eqref{rep-trick} one has to consider $n$ replicas of 
the model ($n$ sheets). The partition function on $n$ independent sheets is ${\mathcal Z}^n=
\int\prod_{k=1}^n{\mathcal D}[\phi^{{}_{(k)}}]e^{-\sum_k {\mathcal S}(\{\phi^{{}_{(k)}}\})}$, where 
$\phi^{{}_{(k)}}$ now denote fields living on the replica $k$. 
We now consider the situation with $n$ coupled sheets. First, on each sheet a branch cut ${\mathcal C}$ 
lying along the spatial direction is introduced (as in Figure~\ref{fig1} (a)). The $n$ replicas are 
coupled through ${\mathcal C}$ (for $n=2$ see Figure~\ref{fig1} (b)). The partition 
function on the $n$ coupled sheets is $\widetilde{\mathcal Z}_n(A)=\int\prod_k{\mathcal D}[\phi^{{}_{
(k)}}]e^{-{\mathcal S}^{(n)}(\{\phi^{{}_{(k)}}\})}$, where  
\begin{multline}
\label{action}
{\mathcal S}^{(n)}=\sum\limits_{k=1}^n\Big\{\sum\limits_{\langle i,j\rangle
\not\perp{\mathcal C}} F(\phi^{{}_{(k)}}_{i},\phi^{{}_{(k)}}_{j})\\
+\sum\limits_{\!\!\!\!\langle i,j\rangle\perp{\mathcal C}}\Big[
J^{(k,k)}F(\phi^{{}_{(k)}}_{i},\phi^{{}_{(k)}}_{j})+J^{(k,k+1)}F(\phi^{{}_{(k)}}_{i},
\phi^{{}_{(k+1)}}_{j})\Big]\Big\}. 
\end{multline}
Here $\langle i,j\rangle\perp {\mathcal C}$ denotes links crossing the cut, and 
$J^{\scriptscriptstyle (k,k')}$ is the coupling between fields next to cut and living on 
replicas $k$ and $k'$. The replicas are coupled in a cyclic fashion, meaning that 
the replica indices $k,k'$ are defined $\textrm{mod}\,n$. The partition function on 
the $n$-sheets Riemann surface ${\mathcal Z}_n(A)$ (cf.~\eqref{rep-trick}) corresponds 
to $J^{\scriptscriptstyle (k,k')}=\delta_{k',k+1}$.

Any ratio $\widetilde{\mathcal Z}_n/{\mathcal Z}^n$ can be calculated using the Jarzynski 
equality~\cite{jarzynski-1997}. Specifically, let us consider a system at equilibrium at 
an initial time $t_i$. Let ${\mathcal Z}_i$ be its partition function. Now let us imagine 
that the system is driven to a new equilibrium state at $t_f$, which is described by 
${\mathcal Z}_f$, using an {\it arbitrary} out-of-equilibrium protocol. For an Hamiltonian 
system this could be a quench, in which some parameters of the Hamiltonian 
${\mathcal H}(t)$ are varied with time. The Jarzynski equality states that 
\begin{equation}
\label{JE}
\Big\langle\exp\Big[-\beta\int_{t_i}^{t_f} dt\delta W(t)\Big]\Big\rangle=
\frac{{\mathcal Z}_f}{{\mathcal Z}_i}.
\end{equation}
Here $\delta W\equiv{\mathcal H}(t+dt)-{\mathcal H}(t)$ is the infinitesimal work performed 
between time $t$ and $t+dt$, $\beta\equiv 1/T$ is the inverse temperature, and $\langle\cdot\rangle$ 
denotes the average over different realizations of the quench protocol. 
The Jarzynski equality has been verified in several 
systems, and it is routinely used to extract free energy differences in out-of-equilibrium 
experiments~\cite{hummer-2001,liphardt-2002,douarche-2005,collin-2005,bustamante-2005,blickle-2006,
harris-2007,saira-2012,jarzynski-2011,seifert-2012,shuoming-2015}, and in Monte Carlo 
simulations (see for instance Ref.~\cite{caselle-2016}). 

The idea of our method is to use~\eqref{JE} performing a quench of the couplings $J^{\scriptscriptstyle 
(k,k')}$ (cf.~\eqref{action}) from the starting configuration with $J^{\scriptscriptstyle (k,k')}=
\delta_{k,k'}$ ($n$ independent replicas) and final one with $J^{\scriptscriptstyle (k,k')}=
\delta_{k',k+1}$ (coupled replicas). Any quench protocol is expected to give the same result. 
We choose the ramp protocol 
\begin{equation}
\label{ramp}
J^{(k,k')}=\left\{
\begin{array}{cc}
\delta_{k,k'} &\textrm{if}\,\, t\le t_i\\
\frac{t-t_i}{t_f-t_i}(\delta_{k',k+1}-\delta_{k,k'})+\delta_{k,k'} &\textrm{if}\,\, t>t_i
\end{array}
\right. 
\end{equation}
Here $t$ is the Monte Carlo time, and $1/(t_f-t_i)\equiv\theta$ is the quenching rate.  
Importantly, $t_i$ has to be large enough to ensure initial thermal equilibrium. 
%
\begin{figure}[t]
\includegraphics*[width=.98\linewidth]{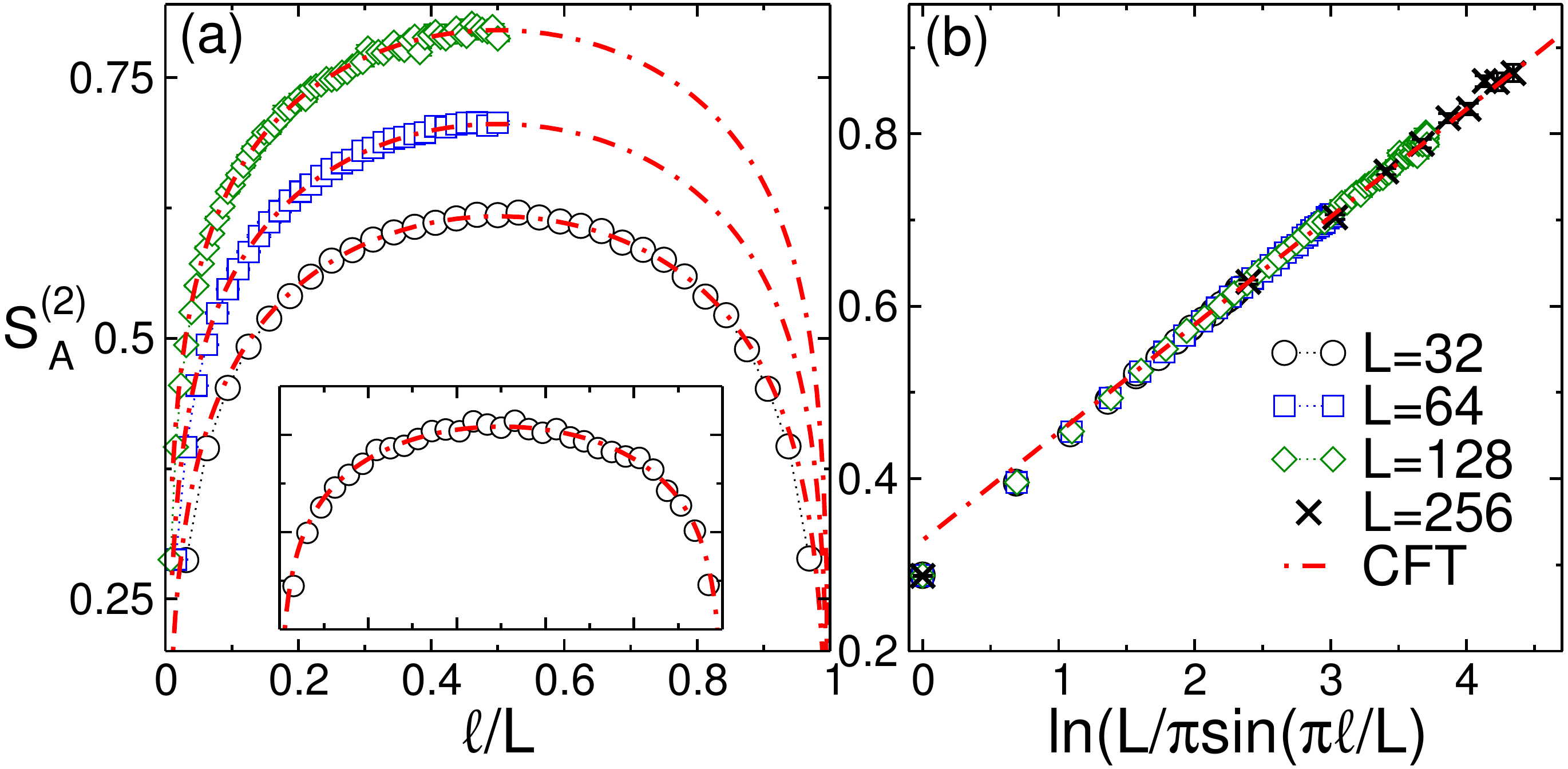}
\caption{ R\'enyi entropy $S^{{}_{(2)}}_A$ in the Ising universality class 
 in $1$$+$$1$ dimensions:  Numerical results using the out-of-equilibrium 
 Monte Carlo method. Panel (a): $S^{{}_{(2)}}_A$ plotted as a function of 
 $\ell/L$, with $\ell$ the length of $A$ and $L=L_x$ (see Figure~\ref{fig1}). 
 Note for $L=32$ the expected symmetry under $\ell\to L-\ell$. The dashed-dotted 
 lines are one-parameter fits to the CFT prediction $S_A^{{}_{(2)}}=
 c/3\log(L/\pi\sin(\pi\ell/L))+k$, with $c=1/2$ the central charge and 
 $k$ a constant.  Inset: $S^{{}_{(2)}}_A$ as obtained from a single 
 simulation. The dashed-dotted line is the same 
 as in the main Figure. Panel (b): Same data as in (a) plotted versus 
 $\log(L/\pi\sin(\pi\ell/L))$. The straight lines are  the same as in (a). 
}
\label{fig2}
\end{figure}
%
Using~\eqref{rep-trick} and~\eqref{JE}, the R\'enyi entropy 
is obtained as 
\begin{equation}
\label{obs}
S_A^{(n)}=\frac{1}{n-1}\ln\big[\langle\exp(-\beta W)\rangle\big],
\end{equation}
where $W$ is the integrated work performed during the Monte Carlo history, i.e., 
$W=(t_f-t_i)^{-1}\sum_{t=t_i}^{t_f}\delta{\mathcal S}^{(n)}(t)$,  
and $\delta {\mathcal S}^{(n)}$ is the change in energy between two 
consecutive Monte Carlo steps $t_i$ and $t_{i+1}$, calculated using the field 
configurations at $t_i$, i.e.,
\begin{equation}
\label{deltaS}
\delta {\mathcal S}^{(n)}=\sum\limits_{k,\langle i,j\rangle\perp{\mathcal C}}
[F(\phi^{{}_{(k)}}_{i},\phi^{{}_{(k+1)}}_{j})-F(\phi^{{}_{(k)}}_{i},\phi^{{}_{(k)}}_{j})]. 
\end{equation}
Importantly, Eq.~\eqref{obs} implies that $S_A^{{}_{(n)}}$ depends on 
the full work distribution function (similar to Ref.~\onlinecite{cardy-2011}). 
Remarkably, in the quasi-static regime one has 
\begin{equation}
\label{ent-work}
S^{(n)}_A=\frac{\beta}{n-1}\Big[-\langle W\rangle+\beta\frac{\sigma_W^2}{2}\Big], 
\end{equation}
where $\sigma_W^2\equiv\langle W^2\rangle-\langle W\rangle^2$ is the work variance. 
Eq.~\eqref{ent-work} is derived assuming that the work distribution function is 
gaussian, and it  is also known as Callen-Welton fluctuation dissipation 
relation~\cite{callen-1951}. The second term in~\eqref{ent-work} corresponds 
to the work $W_d$ dissipated during the quench. Interestingly, Eq.~\eqref{obs} implies 
that the number of independent simulations needed to obtain a reasonable estimate of 
$S_A^{{}_{(n)}}$ increases exponentially with $W_d$ (see Ref.~\onlinecite{jarzynski-2006} 
for a rigorous result). On the other hand from~\eqref{ent-work}, one has that $\sigma_W\to0$ 
in the quasi-static limit $\theta\to\infty$, implying that $S_A^{{}_{(n)}}$ can be 
extracted from a single protocol realization.

\begin{figure}[t]
\includegraphics*[width=0.93\linewidth]{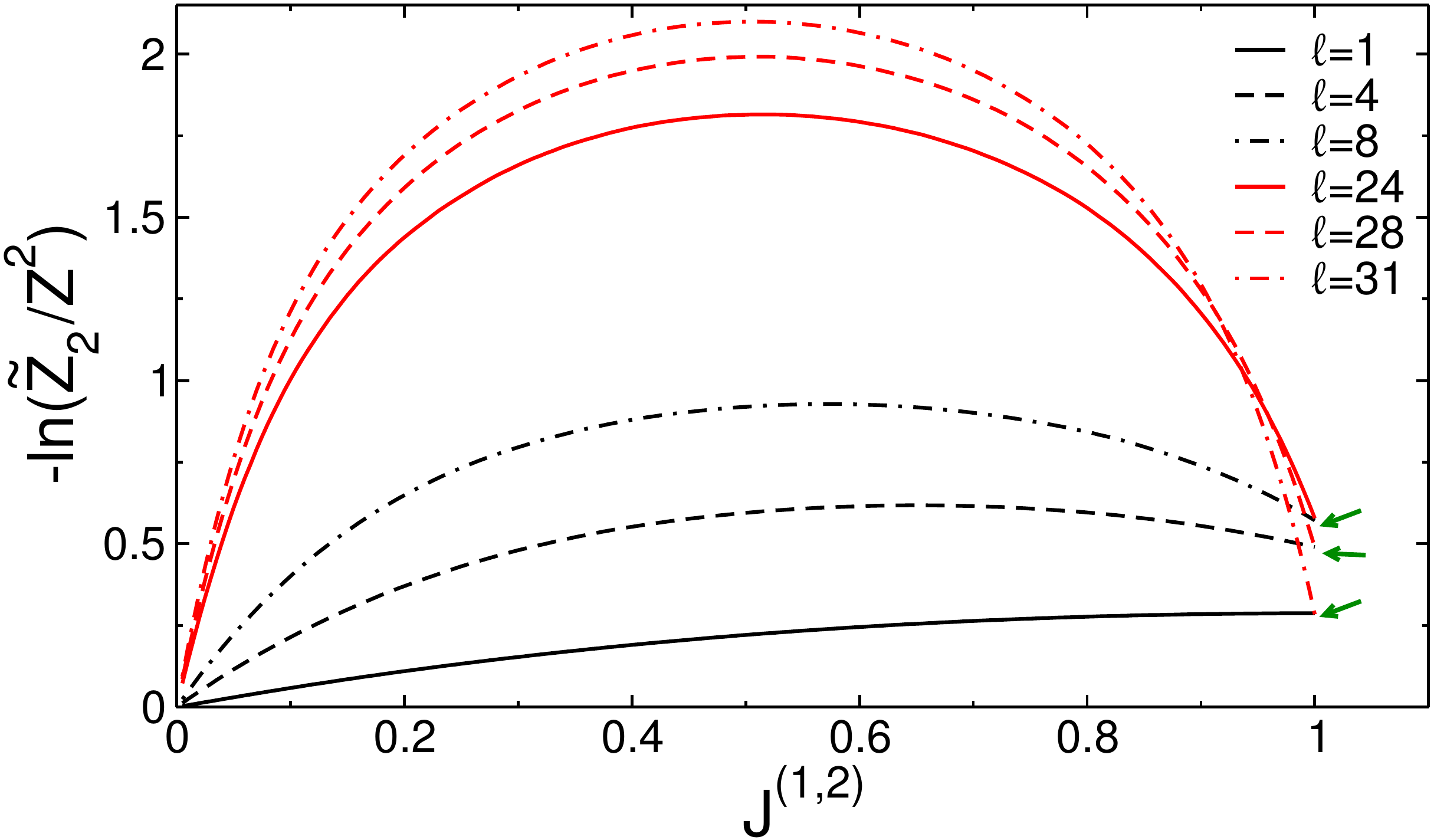}
\caption{Monte Carlo dynamics of the entropy estimator $-\ln({\mathcal Z}_2/
 {\mathcal Z}^2)$. Here $-\ln({\mathcal Z}_2/{\mathcal Z}^2)$ is plotted 
 versus the inter-replicas coupling $J^{\scriptscriptstyle (1,2)}$. 
 Different lines correspond to different sizes $\ell$ of $A$. Data  are for 
 a chain with $L=32$ sites, averaged over $\sim 10$ realizations of the 
 quench. Note the maximum a $J^{\scriptscriptstyle (1,2)}\approx 1/2$. Note 
 also at $J^{\scriptscriptstyle (1,2)}=1$ the expected symmetry under the 
 exchange $\ell\to L-\ell$.  
}
\label{fig3}
\end{figure}

\section{Numerical checks in the Ising universality class}
We now provide numerical evidence supporting the correctness and efficiency of our Monte 
Carlo method. We consider the two-dimensional classical critical Ising model (cf.~\eqref{is-ham}). 
The model has a critical point at 
$\beta_c=\ln(1+\sqrt{2})/2$. Here we consider very elongated lattices with $L_\tau/L_x\gg1$ 
(anisotropic limit). In this limit universal properties are the same as in the one-dimensional 
quantum critical Ising chain at zero temperature, which is defined 
by the Hamiltonian ${\mathcal H}=-\sum_{i=1}^{L_x}(\sigma_i^x\sigma_{i+1}^x+\sigma_i^z)$, 
with $\sigma^{x,z}_i$ the Pauli matrices acting on site $i$ of the chain. The critical 
behavior of both models is described by a Conformal 
Field Theory~\cite{yellow-book} (CFT) with central charge $c=1/2$. 

In our Monte Carlo simulations we employed the Swendsen-Wang method~\cite{swendsen-1987}, 
although any other type of update  
can be used. In the Swendsen-Wang update, at any Monte Carlo time $t$ one assigns to 
each link of the lattice (here the connected $n$ replicas, see Figure~\ref{fig1} (b)) an 
auxiliary activation variable. Links connecting pairs of {\it aligned} spins and not 
crossing the branch cut are then activated with probability $p=1-\exp(-2\beta)$, while links 
connecting aligned spins around the branch cut are activated with probability $p'=1-
\exp(-2\beta J^{\scriptscriptstyle (k,k')})$, with $J^{\scriptscriptstyle (k,k')}$ 
given in~\eqref{ramp}. In our Monte Carlo simulations we fixed $t_i\approx 10^5$ and 
$t_f\approx10^6$. Then, all the different clusters of spins are identified using  
the rule that pairs of spins connected by an activated link are in the same cluster. 
Finally, all the spins belonging to the same cluster are flipped with probability $p=1/2$. 

We should mention that for 
models that can be mapped to the random cluster model, \`a la Fortuin Kasteleyn, it 
is usually convenient to write Monte Carlo observables in terms of cluster-related 
variables. Typically, at least for equilibrium simulations, this leads to a considerable 
speedup. This can be done for the R\'enyi entropy estimator~\eqref{obs}, as we discuss 
in~\ref{clust}. Surprisingly, our results suggest that the cluster estimator performs 
worse than the standard local estimator~\eqref{obs}. This is in contrast with cluster 
estimators for the R\'enyi entropies in equilibrium Monte Carlo simulations, which 
outperform local estimators~\cite{caraglio-2008}. 

\begin{figure}[t]
\includegraphics*[width=0.93\linewidth]{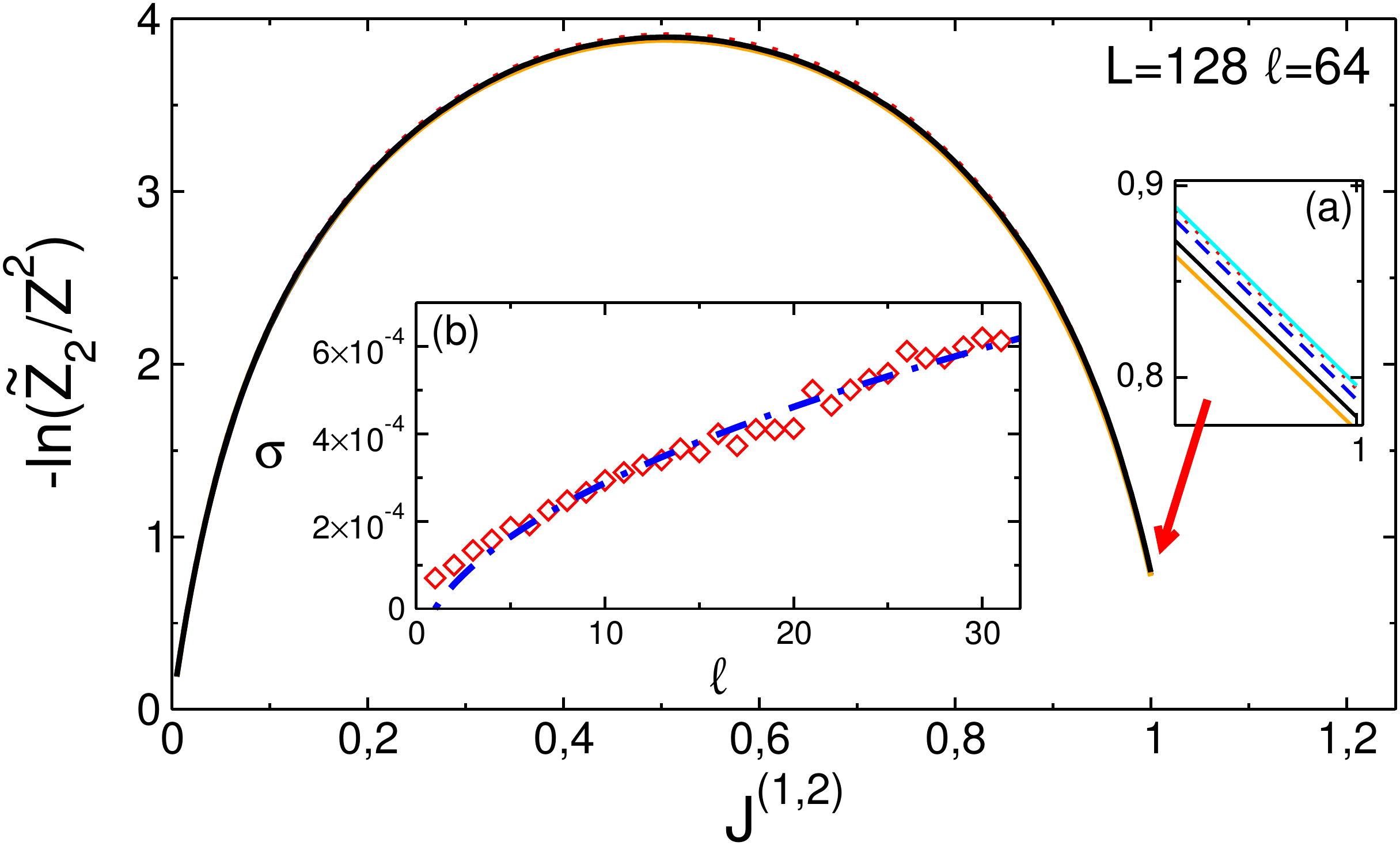}
\caption{ Convergence of the out-of-equilibrium Monte Carlo method 
 for calculating the R\'enyi entropies. $-\ln(\widetilde{\mathcal Z}_2/{\mathcal Z}^2)$ 
 plotted as a function of the inter-replicas coupling $J^{\scriptscriptstyle (1,2)}$. 
 Here $J^{\scriptscriptstyle (1,2)}$ is varied during the Monte Carlo dynamics 
 using~\eqref{ramp}. The different lines correspond to different realizations 
 of the out-of-equilibrium dynamics. Inset (a): Fluctuations of $-\ln(\widetilde{\mathcal Z}_2/
 {\mathcal Z}^2)$ around $J^{\scriptscriptstyle (1,2)}=1$. Inset (b): Standard deviation 
 $\sigma$ of the fluctuations of  $-\ln(\widetilde{\mathcal Z}_2/{\mathcal Z}^2)$ 
 at $J^{\scriptscriptstyle (1,2)}=1$ plotted versus the subsystem length $\ell$. The data are for 
 a chain with $L=32$ sites. The standard deviation is calculated using 
 $\sim 250$ independent realizations of the  out-of-equilibrium protocol. 
 The dashed-dotted line is a fit to the behavior $\propto\ell^{1/2}$. 
}
\label{fig4}
\end{figure}

The results for $S_A^{{}_{(2)}}$ are illustrated in Figure~\ref{fig2}. Panel (a) 
shows $S^{{}_{(2)}}_A$ as obtained using~\eqref{obs} and plotted versus $\ell/L$. The symbols 
are Monte Carlo data for system sizes with $L=32,64,128$, averaged over $\sim 10$ independent 
realizations of the driving protocol. In the scaling limit $\ell,L\to\infty$, the 
behavior of $S_A^{{}_{(n)}}$ is obtained from CFT as~\cite{calabrese-2004} 
\begin{equation}
\label{cft}
S_A^{(n)}=\frac{c}{6}\Big(1+\frac{1}{n}\Big)\ln\Big[\frac{L}{\pi}\sin\Big(\frac{\pi\ell}{L}
\Big)\Big]+k_n,
\end{equation}
where $c=1/2$ is the central charge and $k_n$ a non universal constant. The dashed-dotted lines 
in the Figure are obtained from~\eqref{cft} by fitting  $k_n$, after fixing $c=1/2$. The good 
agreement with the data confirms the validity of the method. This is also clear from the perfect 
linear behavior in panel (b), where we plot $S_A^{{}_{(2)}}$ versus $\ln[L/\pi
\sin(\pi\ell/L)]$. The inset in Figure~\ref{fig2} (a) plots $S_A^{{}_{2}}$ as obtained from a 
single realization of the driving protocol, i.e., a single Monte Carlo simulation. The dashed-dotted 
line is the same as in the main Figure. The good agreement with the data suggests that the 
protocol~\eqref{ramp} with $t_f\approx 10^6$ is already close to the quasi-static regime, at 
least for $L=32$. 

It is also interesting to investigate the behavior of $-\ln(\widetilde{\mathcal Z}_2
/{\mathcal Z}^2)$ as a function of the value of $J^{\scriptscriptstyle (1,2)}$ during the Monte Carlo 
dynamics. This is discussed in Figure~\ref{fig3}. The data are for a single realization of the driving 
protocol. At $J^{\scriptscriptstyle (1,2)}\approx 0$, 
one has $-\log(\widetilde{\mathcal Z}_2/{\mathcal Z}^2)\approx 0$, reflecting that at the early stage 
of the Monte Carlo the two replicas are disconnected. Interestingly, $-\log(\widetilde{\mathcal 
Z}_2/{\mathcal Z}^2)$ exhibits a maximum around $J^{\scriptscriptstyle (1,2)}\approx 1/2$. The expected 
symmetry $S^{{}_{(2)}}_A(\ell)=S_A^{{}_{(2)}}(L-\ell)$, which reflects that the zero-temperature 
one-dimensional system is in a pure state, is observed only at $J^{\scriptscriptstyle (1,2)}=1$ 
(see arrows in the Figure).  

\begin{figure}[t]
\includegraphics*[width=0.93\linewidth]{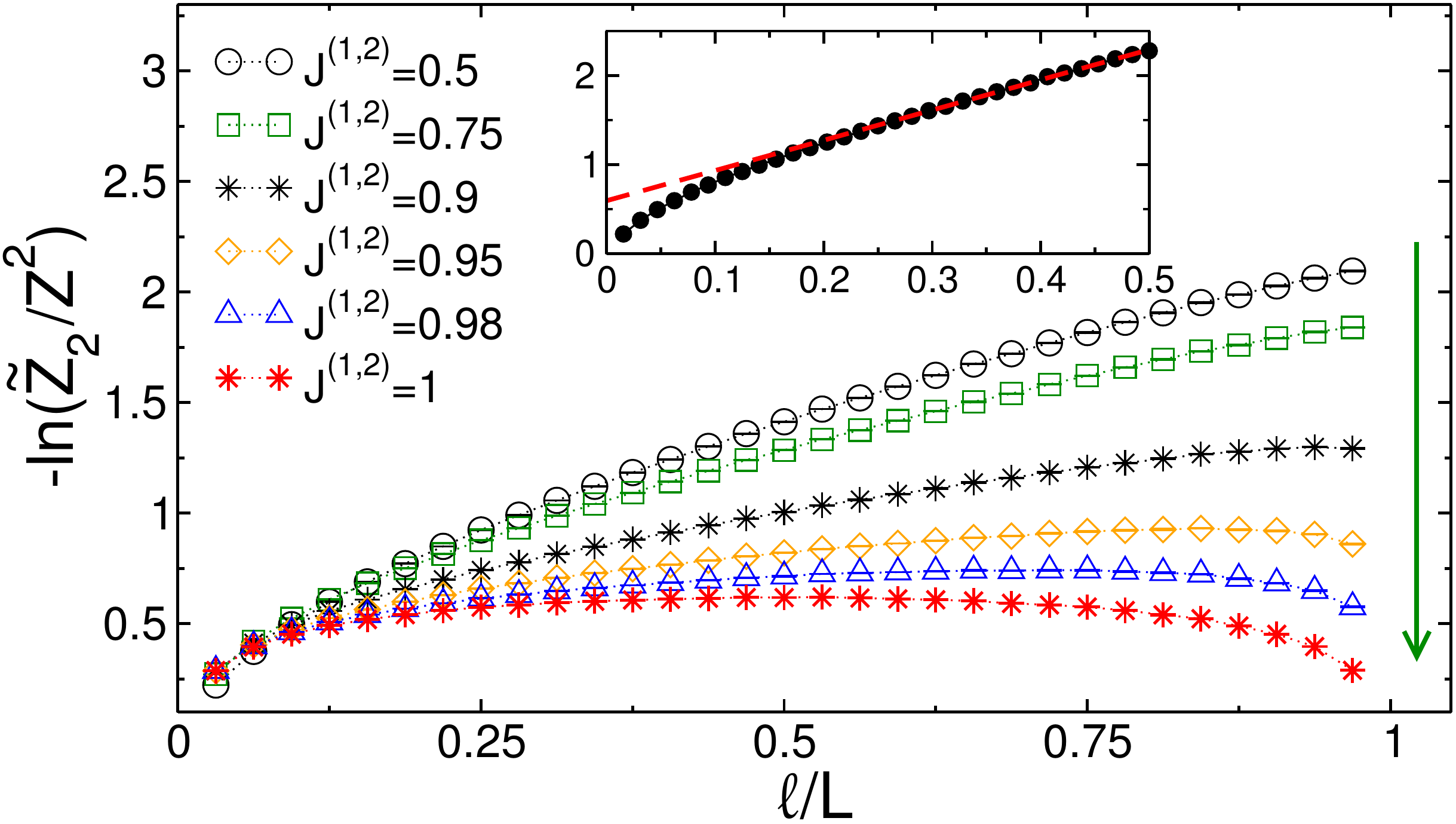}
\caption{ Monte Carlo evolution of the R\'enyi entropy estimator 
 $-\ln(\widetilde{\mathcal Z}_2/{\mathcal Z}^2)$. Here $-\ln(
 \widetilde{\mathcal Z}_2/{\mathcal Z}^2)$ is plotted against $\ell/L$, 
 with $\ell$ and $L$ the subsystem the chain size, respectively. 
 Different symbols correspond to different values of $J^{
 \scriptscriptstyle (1,2)}$. The data are for $L=32$. Inset: 
 The same as in the main figure for $L=64$ at $J^{\scriptstyle (1,2)}=1/2$. 
 The dashed line is a linear fit. 
}
\label{fig5}
\end{figure}

We now discuss the fluctuations of $S_A^{{}_{(2)}}$ between different realizations of the driving protocol. 
Figure~\ref{fig4} plots $-\log(\widetilde{\mathcal Z}_2/{\mathcal Z}^2)$ as a function of $J^{\scriptscriptstyle 
(1,2)}$ for $10$ independent realizations of the driving protocol. The inset (a) shows a zoom around the region 
$J^{\scriptscriptstyle (1,2)}\approx 1$. Fluctuations between different realizations of the dynamics are of the order 
of a few percents. The behavior as a function of $\ell$ of the statistical error $\sigma$ associated 
with the fluctuations between different realizations is illustrated in inset (b). 
Here $\sigma$ is calculated as the standard deviation of the fluctuations, considering a sample of $\sim 250$ 
independent simulations. Clearly, $\sigma$ increases mildly as a function of $\ell$. The dashed-dotted line 
in the inset is a fit to $\propto\ell^{1/2}$. Finally, it is interesting to investigate the dependence 
of $-\log(\widetilde{\mathcal Z}_2/{\mathcal Z}^2)$ on $J^{\scriptscriptstyle (1,2)}$. Figure~\ref{fig5} 
shows $-\log(\widetilde{\mathcal Z}_2/{\mathcal Z}^2)$ as a 
function of $\ell/L$ for several values of $J^{\scriptscriptstyle (1,2)}$. Most notably, for any 
$J^{\scriptscriptstyle (1,2)}<1$ one has that the symmetry $S_A^{{}_{(2)}}(\ell)=S_A^{{}_{(2)}}(L-\ell)$ is 
not present. We should mention that this resembles the behavior of $S_A^{{}_{(n)}}$ when the total system 
is not pure, for instance at finite temperature. Moreover, the data for $J^{\scriptstyle}\ne 1$ suggest 
a volume-law behavior $S_A^{(2)}\propto\ell$ (see inset in Figure~\ref{fig5}), again as in finite temperature systems. 

\section{Conclusions} 

We presented a novel out-of-equilibrium framework for measuring the R\'enyi entropies by combining  
the Jarzynski equality and the replica-trick. As an application, we presented a new classical 
Monte Carlo method to measure the R\'enyi entropies.  

This work opens numerous research directions. First, it would be useful to implement the approach in 
the framework of quantum Monte Carlo. Furthermore, it would interesting to understand 
analytically the behavior of $-\log({\mathcal Z}_2/{\mathcal Z}^2)$ as a function 
of $J^{\scriptscriptstyle (1,2)}$ (see Figure~\ref{fig5}). An intriguing 
possibility is that $J^{\scriptscriptstyle (1,2)}$ could be interpreted as a finite-temperature for 
the one-dimensional system. This could provide an alternative way to obtain the finite-temperature 
R\'enyi entropies. 
Interestingly, our approach could be used for the moments of the partially transposed 
reduced density matrix~\cite{alba-2013,chung-2014}. These are the main ingredients to construct the 
logarithmic negativity~\cite{vidal-2002,plenio-2005,calabrese-2012}, 
which is a good entanglement measure for mixed states. 
Another interesting direction is to consider a protocol in which the length of $A$ is also 
quenched. This would allow to obtain the entropies for different subsystem sizes in a single 
simulation. Finally, it is important to understand whether our framework could provide a viable 
alternative to measure R\'enyi entropies in cold-atom experiments or in NMR quantum 
simulators~\cite{fan-2016,li-2016}.

\paragraph*{Acknowledgments.---} 

I am greatly indebted with Claudio Bonati for drawing to my attention Ref.~\onlinecite{caselle-2016} and 
for useful discussions. I thank Pasquale Calabrese and Paola Ruggiero for reading the manuscript and for 
useful suggestions. I acknowledge support from the ERC under the Starting Grant 279391 EDEQS. This project 
has received funding fromt the European Union's Horizon 2020 research and innovation programme under the 
Marie Sklodowoska-Curie grant agreement No 702612.



\appendix 

\section{Cluster estimator for the R\'enyi entropies} 
\label{clust}

\begin{figure}[t]
\includegraphics*[width=0.98\linewidth]{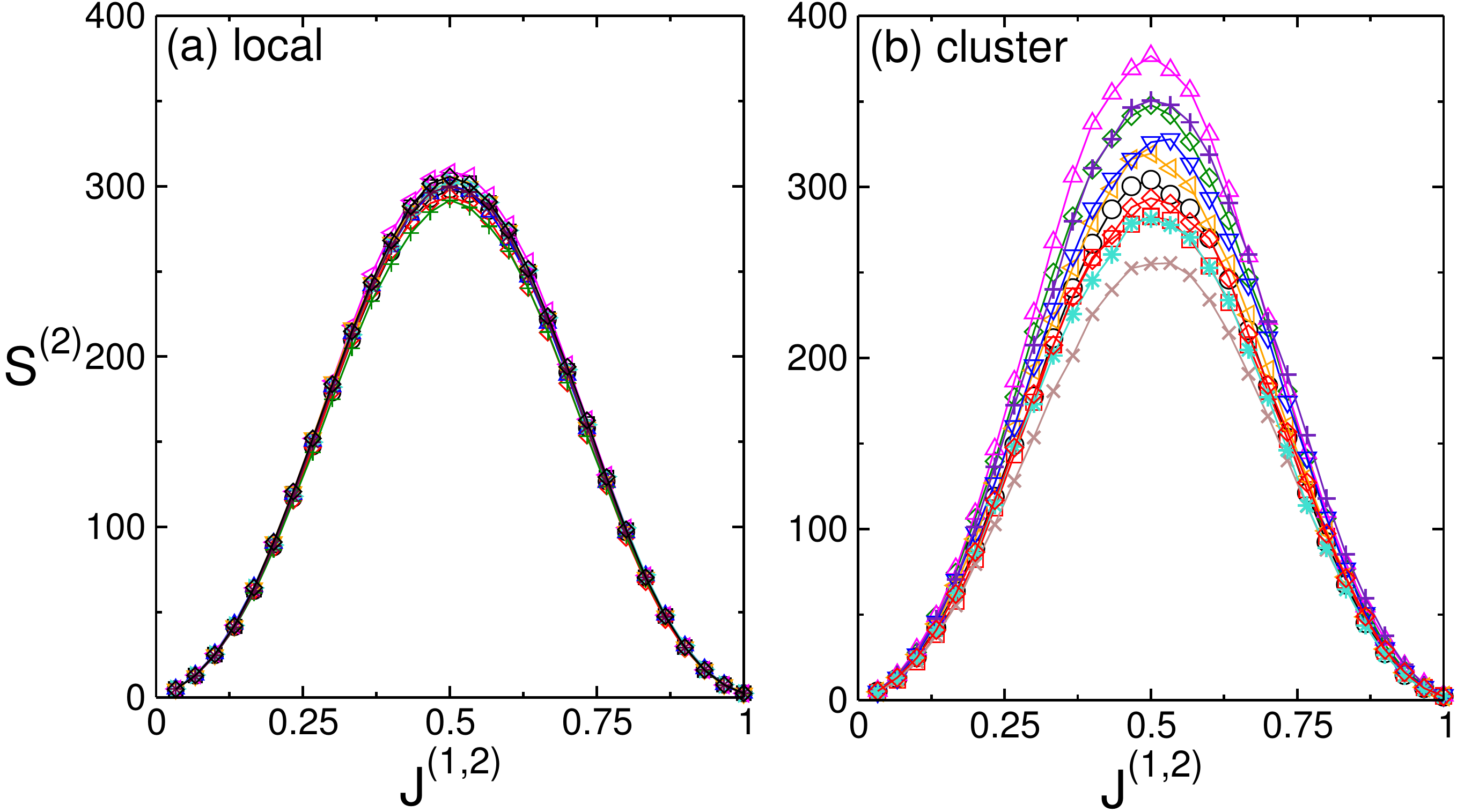}
\caption{ Comparison between the standard estimator for the R\'enyi 
 entropy $S^{(2)}$ (panel (a)) and the cluster-based one (panel (b)). 
 In both panels $S^{(2)}$ is plotted versus the inter-replica coupling 
 $J^{\scriptscriptstyle (1,2)}$. The different symbols correspond to 
 different realizations of the out-of-equilibrium protocol. 
 The data are for a system with $L_y=256$ and subsystem size $\ell=
 101$. Note that the cluster estimator exhibits larger fluctuations 
 between different realizations of the protocol. 
}
\label{fig_app}
\end{figure}

For statistical models that are mappable to the random cluster model, i.e., 
whose partition function admits a representation \'a la Fortuin 
Kasteleyn~\cite{fk-1969}, it is possible to rewrite~\eqref{deltaS} in terms 
of cluster variables (see Ref.~\onlinecite{caraglio-2008} and 
\onlinecite{alba-2013} for similar results in equilibrium Monte Carlo 
simulations).  
Typically, this leads to improved estimators with much smaller variance, as 
compared with the standard one expressed in terms of spin variables. The 
physical reason is that each fixed cluster configuration corresponds to 
many spin configurations. To be concrete, we focus on the two-dimensional 
Ising model on a square lattice, which is defined by the Hamiltonian
\begin{equation}
\label{is-ham}
H=-J\sum\limits_{\langle i,j\rangle} S_iS_j, 
\end{equation}
where $S_i=\pm 1$ and $J$ is the coupling strength. The partition function of 
the Ising model at inverse temperature $\beta$ can be expressed using the 
Fortuin Kasteleyn representation as 
\begin{equation}
\label{fk}
{\mathcal Z}=\sum\limits_{\gamma\in \Omega}\Big\{\prod\limits_{e\in E}
p^{\omega(e)}(1-p)^{1-\omega(e)}\Big\}2^{\kappa(\gamma)}. 
\end{equation}
Here the sum is over the set of subgraphs $\gamma$ on the lattice, i.e., 
arbitrary sets of lattice sites, $\omega(e)=1$ denotes the (activated) 
links connecting points in $\gamma$ (it is $\omega=0$ for the other links), 
while $\kappa(\gamma)$ is the number of connected components of $\gamma$ 
(clusters). In~\eqref{fk}, we defined $p=1-e^{-2\beta J}$. 

To calculate $\textrm{Tr}\rho_A^n$, one first considers a stack of $n$ 
disconnected replicas of the model and of~\eqref{fk}. Each configuration of 
activated links (cluster configuration) on the disconnected sheets is also 
a valid one on the $n$-sheets Riemann surface (see Figure~\ref{fig1}), in 
which the copies are connected. The only difference is that now activated 
links across the branch cut connect sites on different replicas. Importantly, 
the same configuration of activated links corresponds to different numbers of 
clusters on the $n$ independent sheets and the Riemann surface. 
The ratio of partition functions in~\eqref{rep-trick} is written as  
\begin{equation}
\label{c-entr}
\textrm{Tr}\rho_A^n=\frac{{\mathcal Z}_n(A)}{{\mathcal Z}^n}
=\langle 2^{\kappa_{\mathcal C}-\kappa}\rangle_{n}, 
\end{equation}
where, given a fixed set of activated links, $\kappa_{\mathcal C}$ is the total number 
of clusters on the $n$-sheets Riemann surface with cut ${\mathcal C}$, while $\kappa$ is the 
total number of clusters for the model on the $n$ independent sheets. In~\eqref{c-entr} 
the statistical average $\langle\cdot\rangle_{n}$ is performed using the model defined on the $n$ 
independent sheets. Interestingly, from formula~\eqref{c-entr} one has that the entropies 
depend only on the number of clusters. 

In our out-of-equilibrium setup the Monte Carlo dynamics happens on the coupled $n$ sheets 
(see Figure~\ref{fig1}). Clearly, the Fortuin-Kasteleyn representation~\eqref{fk} still 
holds. However, in contrast to~\eqref{fk}, the probability $p$ of activating the links 
across the branch cuts are dynamical variables (cf.~\eqref{ramp}). On the other hand, 
since the change in energy $\delta {\mathcal S}^{(n)}$ (cf.~\eqref{deltaS}) is calculated 
at fixed fields configuration, the number of clusters does not change between two 
consecutive Monte Carlo updates. Thus, one has  
\begin{equation}
\label{c-est}
S_A^{(n)}=\frac{1}{n-1}\ln\Big[\Big\langle\exp\Big(-\sum_{t=t_i}^{t_f}\delta W_c(t)
\Big)\Big\rangle\Big], 
\end{equation}
with 
\begin{equation}
\label{wc}
\delta W_c(t)\equiv \sum\limits_{e\perp{\mathcal C}}\ln\Big(\frac{p(t+\delta t)}{p(t)}
\Big)^{\omega(e)}\Big(\frac{1-p(t+\delta t)}{1-p(t)}\Big)^{1-\omega(e)}. 
\end{equation}
In the definition of $\delta W_c$ the product is over all the links across the branch cut, 
which connect sites both on the same sheet and on different ones, $p(t)=1-\exp(-2
\beta J^{{}_{(k,k')}})$, with $J^{{}_{(k,k')}}$ as defined in~\eqref{ramp}, and $\delta t
=1/(t_f-t_i)$. For small $\delta t$, which is always the case in the simulations, 
from~\eqref{wc} one obtains 
\begin{equation}
\delta W_c(t)=\sum\limits_{e\perp {\mathcal C}}\frac{(\omega(e)-p)\delta p}
{(1-p) p},  
\end{equation}
where $\delta p\equiv p(t+\delta t)-p(t)$. 
It is interesting to observe that~\eqref{c-est} does not depend explicitly on the 
number of clusters, in contrast to~\eqref{c-entr}.

The strategy to use~\eqref{wc} and~\eqref{c-est} in Monte Carlo simulations is straightforward 
using the Swendsen-Wang algorithm. Between two successive update steps of the Monte Carlo one has 
to measure both $p(t)$ and $p(t+\delta t)$, which are then substituted in~\eqref{wc}. Note that 
in~\eqref{wc}, the sum is over the links crossing the branch cuts, and $\omega(e)=1$ only if the 
link $e$ is active. Thus, the infinitesimal work $\delta W_c$ obtained from~\eqref{wc} is 
integrated on the Monte Carlo history, to obtain the exponent in~\eqref{c-est}.

The validity and efficiency of~\eqref{wc} is investigated in Figure~\ref{fig_app}. 
Panels (a) and (b) show the R\'enyi entropy $S^{(2)}$ calculated using~\eqref{obs} 
and~\eqref{wc}, respectively. The data are for the critical Ising model on the lattice 
with $L_y=256$ and $L_x=2560$, and for a subystem with $\ell=101$ sites. In the Figure, 
$S^{(2)}$ is plotted versus the inter-replica coupling $J^{\scriptscriptstyle (1,2)}$. 
In both panels the data are obtained from simulations with $\sim 150000$ Monte Carlo steps. 
The different symbols correspond to different realization of the out-of-equilibrium 
protocol. Surprisingly, the data obtained using the cluter estimator~\eqref{wc} (panel(b)) 
show much larger fluctuations between different realizations as compared with the 
local estimator~\eqref{obs} (panel (a)). This is in costrast with equilibrium Monte Carlo 
simulations for calculating the R\'enyi entropies, where cluster estimators performs 
better than local ones~\cite{caraglio-2008}

\end{document}